\documentclass[preprint,showpacs,groupedaddress,amsmath,amssymb,]{revtex4} 

\usepackage{graphicx}
\usepackage{amsmath}
\usepackage{bm}

\begin{document}

\title{Atomic population distribution in the output ports of cold-atom interferometers with optical splitting and recombination.}

\author{E.~O.~Ilo-Okeke}
\author{Alex~A.~Zozulya}
\affiliation{Department of Physics, Worcester Polytechnic Institute,100 Institute Road, Worcester, Massachusetts 01609, USA} \email[]{
zozulya@wpi.edu}

\date{\today}
\begin{abstract}
Cold-atom interferometers with optical splitting and recombination use off-resonant laser beams to split a cloud of Bose-Einstein condensate
(BEC) into two clouds that travel along different paths and are then recombined again using optical beams. After the recombination, the
BEC in general populates both the cloud at rest and the moving clouds. Measuring relative number of atoms in each of these clouds
yields information about the relative phase shift accumulated by the atoms in the two moving clouds during the interferometric cycle.
We derive the expression for the probability of finding any given number of atoms in each of the clouds,
discuss features of the probability density distribution, analyze its dependence on the relative accumulated phase shift as
a function of the strength of the interatomic interactions, and compare our results with experiment.
\end{abstract}

\pacs{03.75.Dg, 37.25.+k, 03.75.Kk}
\maketitle
%
\section{Introduction}
%
Using wave-like properties of atoms for atomic interferometry has been a subject of intense and extensive study \cite{berman97}. Atoms are sensitive to electromagnetic fields due to their electric and magnetic moments, their mass allows them to be deflected in the gravitational field thereby making them attractive in the measurements of inertial forces \cite{stodolsky79}. Atom interferometers have been used to measure gravitational constant \cite{fixler07}, acceleration \cite{kasevich91,kasevich92,peters01}, electric polarizability \cite{ekstrom95} and fine structure constant \cite{wicht02} to very high accuracy.

The technical realization of neutral-atom interferometers took some time as compared to their electron- and neutron-based counterparts. Part of the reason is that atoms have large mass,
resulting in a smaller de Broglie wavelengths for the same velocity. Also, neutral atoms cannot easily propagate in dense matter unlike, e.~g., neutrons, and therefore require more ingenious ways
to coherently split and diffract the atomic beam. The first atom interferometer \cite{carnal91} realized in a double-slit diffraction geometry, worked with a stream of supersonic gaseous
atoms and used mechanical gratings \cite{carnal91}. Later experiments \cite{giltner95,rasel95} used standing light wave to coherently diffract the atomic beam. The standing light wave
is formed using a laser beam that is detuned from atomic resonance to avoid spontaneous emission, and is retro-reflected by a mirror. The spatially-varying envelope of the standing wave
creates an effective optical potential acting as a diffraction grating for atoms that can be used to split and recombine an atomic beam.

Another technique \cite{kasevich91,moler92} for diffracting an atomic beam exploits Raman transitions between two hyperfine ground states of an atom  via a third quasi-excited state.
The laser pulses (often called Raman pulses), consist of two counterpropagating light beams with frequencies which are different by the Bohr transition
frequency between the two hyperfine states. Absorption of a photon from one laser beam and stimulated re-emission into another one in this case is accompanied by a transition between the
two hyperfine states.

The use of Bose-Einstein Condensates (BECs) \cite{anderson95} in atom interferometers is appealing for many reasons. BEC has narrow momentum distribution that minimizes the spread in
momentum during the splitting and recombination of the atomic cloud and reduces the expansion of the condensate during propagation. Ultracold BEC can be easily manipulated and confined in
a very small area on an atom chip~\cite{wang05}.  Finally, BEC has large coherence length allowing for good fringe contrast and helping
to determine any offset phases more accurately. Since the first experimental demonstration of interference between two different Bose
condensates \cite{andrews97}, several experimental techniques for the manipulation of BECs and different BEC-based interferometric geometries have been proposed and demonstrated \cite{wang05,shin04,collins05,horikoshi06,fattori08}.

In trapped-atom interferometer geometries \cite{shin04,collins05} the BEC is kept in a trap confining the atomic cloud in all three dimensions. This trap is dynamically split into two double-well traps using a standing light wave, to create two arms which were physically separated in space. After some time, the trap is switched off allowing the condensates in each arm to fall,
expand and interfere.

In guided-wave interferometers the BEC is kept in a waveguide. The condensate is tightly confined in two transverse dimensions but allowed to propagate along the third dimension \cite{wang05,garcia06}. The waveguide potential along this guiding dimension is typically weakly parabolic either because of difficulty of completely canceling magnetic field gradients
or by design. A typical example is the Michelson-type single reflection atom interferometer realized in Ref.~\cite{wang05}. In this interferometer, the BEC cloud $\psi_{0}$ is initially at rest in a waveguide.
Splitting pulses consisting of a pair of counterpropagating laser beams detuned from atomic resonance and acting as a diffraction grating, are incident on the cloud. These pulses split the condensate
into two harmonics, $\psi_{+}$  and $\psi_{-}$, moving with the initial velocities $ \pm v_{0}$, respectively. In a single reflection interferometer, the directions of propagation
of these harmonics are reversed at time $T/2$ (where $T$ is the duration of the interferometric cycle), i.e., in the middle of the cycle with the help of a reflection pulse. The harmonics are
then allowed to propagate back and are recombined when they overlap again using the same optical pulses that were used to split the original BEC cloud.  After the recombination, the condensate
is in general in a superposition of  $\psi_{0}$, $\psi_{+}$ and
$\psi_{-}$ with the relative amplitudes depending on the amount of the accumulated phase shift between the arms of the interferometer acquired during the cycle.

In a double reflection interferometer \cite{garcia06,burke08}, the optical reflection pulse is applied twice at times $T/4$ and $3T/4$. After the first reflection pulse, the harmonics change their
direction of propagation and start moving back. They pass through each other, and exchange their positions by the time $3T/4$. The harmonic that was on the right at $T/2$ is now on the left and vice versa.
The second reflection pulse applied at $3T/4$ again reverses the directions of propagation of the harmonics and, finally, they are recombined at time $T$.

Authors of Refs. \cite{burke08,horikoshi07,segal10} investigated interferometric geometry that does not rely not on reflecting optical pulses but instead uses gradient of the
confining waveguide potential for reversing direction of propagation of the BEC harmonics. In this ``free oscillation'' interferometer the moving BEC clouds propagate
in a parabolic confining potential. They slow down as they climb the potential, stop at the their classical turning points after one quarter of the trap period ($T/4$) has elapsed, and turn back.
At $T/2$ the clouds meet at the bottom of the potential, reach again their turning points at $3T/4$ and are recombined at time $T$. The duration of the interferometric cycle is thus
equal to the oscillation period of the parabolic longitudinal waveguide potential $T$.

A Mach-Zehnder-type atom interferometer is another waveguide interferometer that uses BEC \cite{horikoshi06,horikoshi07}. This interferometer shares the same operating principles with the
above-discussed Michelson interferometer. The difference between the two interferometers lies in their splitting techniques. In the Mach-Zehnder-type atom interferometer the two
counterpropagating waves used for a $\pi/2$ splitting pulse are frequency-shifted with respect to another resulting in a traveling optical potential.  This $\pi/2$ pulse transforms
the original BEC at rest into two clouds of equal amplitude. One of these clouds remains at rest and the other propagates with velocity $v_{0}$. A $\pi$ pulse in the middle of the
cycle stops the moving cloud and brings the one that was a rest into motion . Finally, the second $\pi/2$ pulse applied at the end of the cycle recombines the two clouds.

Both in trapped-atom and guided-wave interferometers, the interference fringes depend on the relative phase accumulated by the atomic clouds in different arms during the interferometric cycle.
Apart from the accumulated phase shift induced by fields or interactions of interest during the experiment, unwanted phase may be accumulated due to confinement effects and interatomic interactions resulting in a decrease in visibility of the interference fringes.
For example, in the single-pass interferometer \cite{wang05}, during the propagation the outer edge of each cloud feels a higher potential
than the inner edge (the outer edge is the leading edge in the first half of the cycle when the clouds move away from each other and the trailing edge in the second part when the clouds move toward each other). The outer edge thus accumulates a larger phase than the inner one.
During the recombination, the outer edge of one cloud interferes with the inner edge of another and the phase difference accumulated due to the presence of the confinement potential leads to a coordinate-dependent residual phase across the clouds after the recombination.
Another mechanism for phase accumulation is due to mutual interaction of two BEC clouds $\psi_{-}$ and $\psi_{+}$ when they spatially overlap. During the separation, the inner edge of one cloud interacts with atoms in the other cloud until it has traversed the entire length of the other cloud, while the outer edge of each cloud hardly interacts with any atoms in the other cloud (and similarly during the recombination). As a result, the inner edge accumulates larger phase than the outer edge. The above two contributions have opposite signs but different magnitudes so the net coordinate-dependent phase is not zero.
Still another mechanism of accumulation of the unwanted phase is due to the fact that the velocities $\pm v$ of the moving harmonics $\psi_{+}$ and $\psi_{-}$ during the reflection are different from their initial velocities $\pm v_{0}$ due to the influence of the confining potential and the interatomic interactions. As as result, the reflection pulses (which are formed by the same pair of counterpropagating beams as the splitting/recombination pulses) are not exactly on resonance and do not exactly reverse the clouds' velocities (the direction of propagation of each of the clouds does change but the speed before and the after the reflections is different).

Theoretical analysis of the single- and double-reflection interferometer geometries has been carried out in Refs.~\cite{olshanii05,stickney07,burke08,stickney08}. According to the analysis of Refs.~\cite{burke08,stickney08}, symmetric motion of the two clouds in the double-reflection geometry partially cancels the velocity errors imposed by the reflection pulses and
the phase imposed by the confining potential \cite{burke08,stickney08}. This conclusion has been confirmed experimentally in Ref.~\cite{burke08}. The ``free oscillation'' interferometer
provides even greater degree of cancelation of the unwanted coordinate-dependent phase since it does not rely on reflection pulses and does not suffer from the velocity mismatch effects.
Recent experiments \cite{horikoshi07,burke08} where the atomic clouds were allowed to be reflected from their classical turning points instead of using reflection pulses to truncate
their motion, confirmed more accurate cancelation of the unwanted phase.

An additional mechanism that could lead to dephasing of the interference fringes is due to atom-atom interactions within each of the two clouds.
These interactions result in the so-called phase diffusion in the BEC clouds \cite{castin97,javanainen97,wright97,jo07,li07}.
The aim of the present paper is to analyze effects of the atom-atom interactions within each of the BEC clouds on the operation of cold-atom
interferometers using optical pulses for splitting and recombination of the condensate. The basis physics of the interferometric cycle in the presence of the interatomic
interactions can be described as follows: The BEC which is initially in a number state with $N$ atoms is split by the optical pulses into two clouds moving in
opposite directions. After the splitting the system is in an entangled state with each cloud being a linear superposition of number states. This superposition is
concentrated around the mean $N/2$ with the relative uncertainty of the order of $1/\sqrt{N}$. In the presence of interactomic interactions, each number
state evolves with different rate resulting in the accumulation of relative time-dependent phases between the different number states. The recombination process
is sensitive to these phases and thus the interactions should influence the contrast of the interferometric fringes. Specifically, we shall derive the expression for the
probability density of observing any given number of atoms in each of the three output ports of the optical beamsplitter/recombiner and analyze it in different limits.
Both the splitting and the recombination (detection) of the BEC are treated rigorously due to a large difference between the characteristic momenta of the condensate and
the momentum imparted to the atoms by the optical beams.  Thus our analysis does not rely on any ad hoc assumptions about the process of detection and can be considered
an analysis from the first principles. The previously-discussed effects resulting in a coordinate-dependent phase due to confinement and the interaction
between the overlapping clouds are not accounted for by the present analysis, i.e., we assume that most of the time the clouds are spatially separated.

The remainder of the paper is organized as follows. In Sec.~\ref{sec:beamsplitter}, we discuss operation of the optical beamsplitter/recombiner and obtain the
expression for the state vector of the system after recombination. In Sec.~\ref{sec:probability}, we derive the expression for the probability density.
Characteristic features of the probability density including the mean, the standard deviation etc., are calculated and the discussed in Sec.~\ref{sec:features}.
The results are compared with the experiment in Sec.~\ref{sec:discussion}.

\section{\label{sec:beamsplitter}Optical beamsplitter operation}
%
Consider BEC cloud $\psi_{0}$ at rest in a confining potential before the beginning of the interferometric cycle.
As was discussed in the introduction, the splitting optical pulses transform the initial BEC cloud into two clouds $\psi_{\pm 1}$
moving in the opposite directions. The clouds are allowed to evolve during the time $T$ and are subject to the action of the recombination optical pulses
(which are identical to splitting optical pulses).
After the recombination, the atoms in general populate all three clouds $\psi_{0}$ and $\psi_{\pm}$. The relative population of
the clouds depends on the phase difference between the clouds $\psi_{\pm}$ acquired during the interferometric cycle.

Let $b^{\dagger}_{0}$, $b^{\dagger}_{-1}$ and $b^{\dagger}_{+1}$  be operators which, acting on a vacuum state, create
an atom belonging to a cloud at rest and moving to the left or right, respectively.

The many-body Hamiltonian describing atomic BEC during the interferometric cycle is of the form
\begin{eqnarray}\label{field_operator_hamiltonian}
    H = \int d {\bf r} \hat{\Psi}^{\dagger}({\bf r},t)H_{0}\hat{\Psi}({\bf r},t)
    + \frac{U_{0}}{2}\int d{\bf r} \hat{\Psi}^{\dagger}({\bf r},t)\hat{\Psi}^{\dagger}({\bf r},t)
    \hat{\Psi}({\bf r},t)\hat{\Psi}({\bf r},t),
\end{eqnarray}
where $H_{0}$ is a single-particle Hamiltonian and $U_{0} = 4\pi\hbar^{2}a_{sc}/M$,
with $M$ being the atomic mass and  $a_{sc}$ the s-wave scattering length. The single-particle Hamiltonian $H_{0}$ accounts for the confining
potential for the BEC and also includes effects of the environment resulting in different rates of evolution for the BEC clouds propagating
in opposite directions. Finally, the quantity $\hat{\Psi}({\bf
r},t)$ is the field operator
\begin{equation}\label{twomodeanzatz}
    \hat{\Psi}({\bf r},t) = b_{+1}\psi_{1}({\bf r},t) + b_{-1}\psi_{-1}({\bf r},t),
\end{equation}
where $\psi_{\pm 1}({\bf r},t)$ are wave functions of the BEC clouds moving to the right and left, respectively. The functions $\psi_{\mp 1}$ are
shifted versions of each other propagating in opposite directions.

Using Eq.~(\ref{twomodeanzatz}) in (\ref{field_operator_hamiltonian}) results in the following Hamiltonian describing effects of the environment and the
interactomic interactions:
\begin{equation}\label{hamiltonian_bbdagger}
    H_{eff} = - \frac{W}{2}\left(\hat{n}_{+1} - \hat{n}_{-1}\right) + g\left(\hat{n}_{+1}^{2} + \hat{n}_{-1}^{2}\right).
\end{equation}
Here $\hat{n}_{\pm 1} = b_{\pm 1}^{\dagger}b_{\pm 1}$ are the number operators, $W$ is the relative environment-introduced energy shift between the right- and left-propagating clouds,
and
\begin{equation}\label{g_definition}
    g = \frac{U_{0}}{2}\int d\bm{r}|\psi_{+1}|^{4} = \frac{U_{0}}{2}\int d\bm{r}|\psi_{-1}|^{4}
\end{equation}
is the coefficient characterizing strength of the interatomic interaction. The Hamiltonian (\ref{hamiltonian_bbdagger}) neglects effects due to
overlap of the right- and left-propagating clouds assuming that most of the time the clouds are spatially separated.

The state vector of the system at the beginning of the interferometric cycle before the splitting pulses is given by the relation
\begin{equation}\label{psi_initial}
    |\Psi_{ini}\rangle = \frac{\left(b^{\dagger}_{0}\right)^{N}}{\sqrt{N!}}|0\rangle,
\end{equation}
where $|0\rangle$ is the vacuum state and $N$ is the total number of atoms in the BEC.

Splitting and recombination pulses couple the operators $b_{\pm 1}$ and $b_{0}$
according to the rules (see Appendix in Ref.~\cite{stickney07} for the corresponding splitting/recombination matrices):
\begin{eqnarray}\label{splitting_pulses_rules}
    &&b_{-1} \rightarrow
    \frac{1}{2}b_{+1} + \frac{e^{-i\beta}}{\sqrt{2}}b_{0} - \frac{1}{2}b_{-1}, \nonumber \\
    &&b_{0} \rightarrow \left(b_{+1} + b_{-1}\right)\frac{e^{-i\beta}}{\sqrt{2}}, \nonumber \\
    &&b_{+1} \rightarrow
    -\frac{1}{2}b_{+1} + \frac{e^{-i\beta}}{\sqrt{2}}b_{0} + \frac{1}{2}b_{-1},
\end{eqnarray}
where $\beta$ is a phase factor.

A single-atom state is transformed by the splitting pulses as
\[b_{0}^{\dagger}|0\rangle \rightarrow \frac{1}{\sqrt{2}}(b_{+1}^{\dagger} + b_{-1}^{\dagger})|0\rangle,\]

so the product state vector of the $N-$ particle system Eq.~(\ref{psi_initial}) after the splitting acquires the form
\begin{equation}\label{psi_after_splitting}
   |\Psi_{split}\rangle = \frac{1}{\sqrt{2^{N}N!}}\left(b^{\dagger}_{+1} + b^{\dagger}_{-1}\right)^{N}|0\rangle = \frac{1}{\sqrt{2^{N}}}\sum_{n=0}^{N}\binom{N}{n}^{1/2}|n,N-n\rangle,
\end{equation}
where
\begin{equation}\label{np_nm_state}
    |n_{+},n_{-}\rangle = \frac{\left(b_{+1}^{\dagger}\right)^{n_{+}}}{\sqrt{n_{+}!}}\frac{\left(b_{-1}^{\dagger}\right)^{n_{-}}}{\sqrt{n_{-}!}}|0\rangle
\end{equation}
is the state with $n_{+}$ atoms traveling to the right and $n_{-}$ to the left, respectively, and $\binom{N}{n} = N!/n!(N-n)!$ is the binomial coefficient.

Time evolution of the state vector is governed by the Hamiltonian (\ref{hamiltonian_bbdagger}):
\[|\Psi(t)\rangle = |\Psi_{split}\rangle \exp\left[-(i/\hbar)\int^{t}_{0}H_{eff}dt^{\prime}\right].\]
States $|n_{+},n_{-}\rangle$ given by Eq.~(\ref{np_nm_state}) are eigenstates of the Hamiltonian (\ref{hamiltonian_bbdagger})  with the eigenvalues
\begin{equation}
    E(n_{+},n_{-}) = - \frac{W}{2}(n_{+} - n_{-}) + g(n_{+}^{2} + n_{-}^{2}).
\end{equation}
The state vector of the system at time $T$ immediately before the recombination is thus given by the relation
\begin{eqnarray}\label{psi_before_recombination}
    &&|\Psi(T)\rangle = \frac{1}{\sqrt{2^{N}N!}}\sum_{n=0}^{N}\binom{N}{n}e^{i \theta(n-N/2) - i\xi[n^{2} + (N-n)^{2}]} \nonumber \\
    &&\times \left(b_{+1}^{\dagger}\right)^{n}\left(b_{-1}^{\dagger}\right)^{N-n}|0\rangle.
\end{eqnarray}
Here
\begin{equation}
    \theta = \frac{1}{\hbar}\int_{0}^{T} dt W
\end{equation}
is the environment-introduced accumulated phase difference between the right and the left clouds and
\begin{equation}\label{xi_definition}
    \xi = \frac{1}{\hbar}\int_{0}^{T} dt g
\end{equation}
is the accumulated nonlinear phase per atom due to interatomic interactions.
The recombination pulses act on $|\Psi(T)\rangle$ in accordance with Eq.~(\ref{splitting_pulses_rules}). The resulting state vector of the system after the recombination
has the form:
\begin{eqnarray}\label{psi_after_recombination}
   &&|\Psi_{rec}\rangle = \frac{1}{\sqrt{2^{N}N!}}\sum_{n=0}^{N}\binom{N}{n}e^{ i\theta(n-N/2) - 2i\xi(n - N/2)^{2}} \nonumber \\
   &&\times \left(-\frac{1}{2}b^{\dagger}_{+1} + \frac{e^{i\beta}}{\sqrt{2}}b^{\dagger}_{0} + \frac{1}{2}b^{\dagger}_{-1}\right)^{n}
   \left(\frac{1}{2}b^{\dagger}_{+1} + \frac{e^{i\beta}}{\sqrt{2}}b^{\dagger}_{0} - \frac{1}{2}b^{\dagger}_{-1}\right)^{N-n}|0\rangle
\end{eqnarray}
where we have omitted irrelevant phase term $\exp(-i\xi N^{2}/2)$.
%
\section{\label{sec:probability}Probability density}
%
The state with $n_{+}$ atoms being in the cloud moving to the right, $n_{-}$ in the cloud moving to left and
$n_{0} = N - n_{+} - n_{-}$ in the cloud at rest, is described by the state vector
\begin{equation}\label{eigenstate_np_nm_no}
    |n_{+},n_{-},n_{0}\rangle = \frac{\left(b^{\dagger}_{+1}\right)^{n_{+}}}{\sqrt{n_{+}!}}\frac{\left(b^{\dagger}_{-1}\right)^{n_{-}}}{\sqrt{n_{-}!}}
    \frac{\left(b^{\dagger}_{0}\right)^{n_{0}}}{\sqrt{n_{0}!}}|0\rangle
\end{equation}
The probability of this outcome after the recombination is given by the modulus squared of the probability amplitude $\langle n_{+},n_{-},n_{0}|\Psi_{rec}\rangle$.
Using Eq.~(\ref{psi_after_recombination}), this probability amplitude can be written as
\begin{eqnarray}\label{probamplitude_ini}
    &&\langle n_{+},n_{-},n_{0}|\Psi_{rec}\rangle = \frac{1}{\sqrt{2^{N}N!}}\sum_{n=0}^{N}\binom{N}{n}e^{i\theta(n-N/2) - 2i\xi(n-N/2)^{2}} \nonumber \\
    &&\langle 0|\frac{\left(b_{0}\right)^{n_{0}}}{\sqrt{n_{0}!}}\frac{\left(b_{-1}\right)^{n_{-}}}{\sqrt{n_{-}!}} \frac{\left(b_{+1}\right)^{n_{+}}}{\sqrt{n_{+}!}} \nonumber \\
    &&\left(-\frac{b^{\dagger}_{+1} - b^{\dagger}_{-1}}{2} + \frac{e^{i\beta}}{\sqrt{2}}b^{\dagger}_{0}\right)^{n}
   \left( \frac{b^{\dagger}_{+1} - b^{\dagger}_{-1}}{2} + \frac{e^{i\beta}}{\sqrt{2}}b^{\dagger}_{0}\right)^{N-n}|0\rangle.
\end{eqnarray}
Equation~(\ref{probamplitude_ini}) can be recast as
\begin{eqnarray}\label{probamplitude_tmp1}
    &&\langle n_{+},n_{-},n_{0}|\Psi_{rec}\rangle = \frac{1}{\sqrt{2^{N}N!}}\sum_{n=0}^{N}\binom{N}{n}e^{i\theta (n - N/2) - 2i\xi(n-N/2)^{2}} \nonumber \\
    &&\frac{1}{\sqrt{n_{+}!n_{-}!n_{0}!}}\langle 0|\frac{\partial^{n_{0}}}{\partial (b_{0}^{\dagger})^{n_{0}}}\frac{\partial^{n_{-}}}{\partial (b_{-1}^{\dagger})^{n_{-}}}\frac{\partial^{n_{+}}}{\partial (b_{+1}^{\dagger})^{n_{+}}} \nonumber \\
    &&\left(-\frac{b^{\dagger}_{+1} - b^{\dagger}_{-1}}{2} + \frac{e^{i\beta}}{\sqrt{2}}b^{\dagger}_{0}\right)^{n}
   \left(\frac{b^{\dagger}_{+1} - b^{\dagger}_{-1}}{2} + \frac{e^{i\beta}}{\sqrt{2}}b^{\dagger}_{0}\right)^{N-n}|0\rangle.
\end{eqnarray}
The product of two terms in parentheses can be represented as the double sum
\begin{eqnarray}
    &&\left(-\frac{b^{\dagger}_{+1} - b^{\dagger}_{-1}}{2} + \frac{e^{i\beta}}{\sqrt{2}}b^{\dagger}_{0}\right)^{n}
   \left(\frac{b^{\dagger}_{+1} - b^{\dagger}_{-1}}{2} + \frac{e^{i\beta}}{\sqrt{2}}b^{\dagger}_{0}\right)^{N-n} \nonumber \\
   &&= \sum_{i=0}^{n}\sum_{j=0}^{N-n}\binom{n}{i}\binom{N-n}{j}\left(\frac{e^{i\beta}b_{0}^{\dagger}}{\sqrt{2}}\right)^{i+j}(-1)^{n-i}\left(\frac{b_{+1}^{\dagger} - b_{-1}^{\dagger}}{2}\right)^{N-i-j}.
\end{eqnarray}
The derivatives with respect to $b_{0}$ in Eq.~(\ref{probamplitude_tmp1}) will select only the term with $i+j = n_{0}$ from this sum, yielding
\begin{eqnarray}\label{probamplitude_nonzero_phi}
    \langle n_{+},n_{-},n_{0}|\Psi_{rec}\rangle = e^{i\alpha}\sqrt{\frac{N!2^{n_{0}-N}}{n_{0}!n_{+}!n_{-}!}}\sum_{n=0}^{N}e^{i \theta(n-N/2) - 2i\xi(n-N/2)^{2}}I(n,n_{0})
\end{eqnarray}
where
\begin{equation}\label{Inn0_ini}
    I(n,n_{0}) = i^{N-n_{0}}\frac{\binom{N}{n}\binom{N}{n_{0}}^{-1}}{2^{N}}\sum_{i=0}^{n}\sum_{j=0}^{N-n}\delta_{i+j,n_{0}}\binom{n}{j}\binom{N-n}{j}(-1)^{n - i}.
\end{equation}
and
\begin{equation}\label{phase_alpha}
    \alpha = \beta n_{0} + (\pi/2)(n_{-} - n_{+}).
\end{equation}
The unwieldy expression (\ref{Inn0_ini}) can be written in a much more manageable form by evaluating Eq.~(\ref{probamplitude_tmp1}) for $\xi = 0$.
In this case, summation in Eq.~(\ref{probamplitude_tmp1}) can be readily carried out and, after differentiation, Eq.~(\ref{probamplitude_tmp1})
results in the expression
\begin{eqnarray}\label{probamplitude_zero_phi_fin}
    \langle n_{+},n_{-},n_{0}|\Psi_{rec}\rangle = e^{i\alpha}\sqrt{\frac{N!2^{(n_{0} - N)}}{n_{0}!n_{+}!n_{-}!}}\left(\cos\frac{\theta}{2}\right)^{n_{0}}\left(\sin\frac{\theta}{2}\right)^{N - n_{0}}.
\end{eqnarray}
Comparison of Eq.~(\ref{probamplitude_nonzero_phi}) and Eq.~(\ref{probamplitude_zero_phi_fin}) shows that
\begin{equation}
    \sum_{n=0}^{N}e^{i n \theta}I(n,n_{0}) = e^{i\theta N/2}\left(\cos\frac{\theta}{2}\right)^{n_{0}}\left(\sin\frac{\theta}{2}\right)^{N-n_{0}}
\end{equation}
immediately yielding
\begin{equation}\label{Inn0}
    I(n,n_{0})  = \frac{1}{\pi}\int_{0}^{\pi}dx e^{ix(N - 2n)}\left(\cos x\right)^{n_{0}}\left(\sin x\right)^{N-n_{0}},
\end{equation}
Using Eq.~(\ref{probamplitude_nonzero_phi}), we can write the probability density
\[P(n_{+},n_{-},n_{0}) = |\langle n_{+},n_{-},n_{0}|\Psi_{rec}\rangle|^{2}\]
as the product of two functions:
\begin{equation}\label{prob_dens_definition}
    P(n_{+},n_{-},n_{0}) = P_{\pm}(n_{+},n_{-},n_{0})P_{0}(n_{0},\theta,\xi),
\end{equation}
where
\begin{equation}\label{Ppm}
    P_{\pm} = \frac{(N-n_{0})!}{n_{+}!n_{-}!2^{N-n_{0}}}
\end{equation}
and
\begin{equation}\label{Pn0}
    P_{0}(n_{0},\theta,\xi) = \binom{N}{n_{0}}\left|\Sigma(n_{0},\theta,\xi)\right|^{2}.
\end{equation}
with the function $\Sigma(n_{0}\theta,\xi)$ given by the relation
\begin{equation}\label{Sigma}
    \Sigma(n_{0},\theta,\xi) = \sum_{n=0}^{N}e^{i\theta(n-N/2) - 2i\xi(n-N/2)^{2}}I(n,n_{0}).
\end{equation}
The function $P_{\pm}$ describes the probability of observing $n_{+}$ and $n_{-}$ atoms in the right and left clouds, respectively, for any given number $n_{0} = N - (n_{+} + n_{-})$
atoms in the central cloud. This function is independent both on $\theta$ and the nonlinearity $\xi$ and is normalized to one:
\begin{equation}\label{Ppm_normalization}
    \sum_{n_{+} = 0}^{N-n_{0}}P_{\pm} = \sum_{n_{+} = 0}^{N-n_{0}}\frac{(N-n_{0})!2^{n_{0} - N}}{n_{+}!(N-n_{0} - n_{+})!} = 1.
\end{equation}
With the use of Stirling's approximation in Eq.~(\ref{Ppm}), $P_{\pm}$ can be simplified to:
\begin{equation}\label{Ppm_Stirling}
    P_{\pm} = \sqrt{\frac{2}{\pi(N-n_{0})}}\exp\left[\frac{(n_{+} - n_{-})^{2}}{2(N-n_{0})}\right], \; n_{+} + n_{-} = N-n_{0}.
\end{equation}
The function $P_{0}(n_{0},\theta,\xi)$ describes the probability of observing $n_{0}$ atoms in the central cloud.
The effects of both the external phase $\theta$ and the nonlinearity $\xi$ are contained in this function. It is also normalized to one:
\begin{equation}\label{Pn0_normalization}
    \sum_{n_{0} = 0}^{N}P_{0}(n_{0},\theta,\xi) = 1.
\end{equation}
The function $\Sigma$ (\ref{Sigma}) satisfies the symmetry relations
\begin{eqnarray}\label{Sigma_relations}
    &&\Sigma(n_{0},-\theta) = (-1)^{N-n_{0}}\Sigma(n_{0},\theta),\nonumber \\
    &&\Sigma(n_{0},\pi -\theta) = \Sigma(N-n_{0},\theta).
\end{eqnarray}
The probability function $P_{0}$ given by Eq.~(\ref{Pn0}) is periodic in $\theta$ with the period $2\pi$.
Relations (\ref{Sigma_relations}) alow us in the following to restrict our analysis to the values of
$\theta$ lying in the interval $0 \le \theta \le \pi/2$ since
\begin{eqnarray}\label{Pn0_relations}
    &&P_{0}(n_{0},-\theta) = P_{0}(n_{0},\theta), \nonumber \\
    &&P_{0}(n_{0},\pi -\theta) = P_{0}(N-n_{0},\theta).
\end{eqnarray}
%
\section{\label{sec:evaluation}Evaluating the probability density function $P_{0}$}
%
The exact expression for the probability density distribution function $P_{0}$ given by the equations (\ref{Pn0}), (\ref{Inn0}) and (\ref{Sigma}),
is relatively complex and does not lend itself readily to an easy interpretation.
In the following we shall transform and simplify it to make it more amenable for the subsequent analysis.

The integral $I(n,n_{0})$ given by Eq.~(\ref{Inn0}) can be evaluated in the complex plane by the method of steepest-descent to yield:
\begin{eqnarray}\label{estimate_I}
    &&I(n,n_{0}) = \frac{1}{\sqrt{\pi N}}\exp\left[(N-n_{0})\ln\sqrt{1 - \frac{n_{0}}{N}} + n_{0}\ln \sqrt{\frac{n_{0}}{N}}-\frac{(n-N/2)^{2}}{N}\right]\nonumber \\
    &&\times \left[e^{i(N-2n)\arccos\sqrt{n_{0}/N}} + (-1)^{N-n_{0}}e^{-i(N-2n)\arccos\sqrt{n_{0}/N}}\right].
\end{eqnarray}
Using Eq.~(\ref{estimate_I}) in the expression for $\Sigma$ (\ref{Sigma}), approximating summation by integration and evaluating the integral, we get:
\begin{eqnarray}\label{Sigma_estimate}
    &&\Sigma(n_{0}) = \frac{1}{\sqrt{1 + 2i\xi N}}\exp\left[(N-n_{0})\ln\sqrt{1 - \frac{n_{0}}{N}} + n_{0}\ln \sqrt{\frac{n_{0}}{N}}\right] \nonumber \\
    &&\times \left[e^{-\eta_{-}} + (-1)^{N-n_{0}}e^{-\eta_{+}}\right],
\end{eqnarray}
where
\begin{equation}\label{P0_exp_arguments}
    \eta_{\mp} = \frac{N\left(\arccos\sqrt{n_{0}/N} \mp \theta/2\right)^{2}}{1 + 2i\xi N}.
\end{equation}
Finally, using the Stirling's approximation and Eq.~(\ref{Sigma_estimate}) in Eq.~(\ref{Pn0}) results in the expression for the probability density $P_{0}$:
\begin{eqnarray}\label{Pnn0_analytical}
    P_{0}(n_{0}) = \frac{1}{\sqrt{2\pi(1 + 4\xi^{2}N^{2})}}\sqrt{\frac{N}{n_{0}(N-n_{0})}}\left|e^{- \eta_{-}} + (-1)^{N-n_{0}}e^{-\eta_{+}}\right|^{2}.
\end{eqnarray}
Equation~(\ref{Pnn0_analytical}) is not applicable at the two end points, $n_{0}$ and $n_{0} = N$, where it has to be replaced by the expressions
\begin{eqnarray}\label{Pnn0_analytical_endpoints}
    &&P_{0}(0) = \frac{1}{\sqrt{1 + \xi^{2} N^{2}}}\exp\left[\frac{-2(\pi - \theta)^{2}N}{1 + \xi^{2} N^{2}}\right],\nonumber \\
    &&P_{0}(N) = \frac{1}{\sqrt{1 + \xi^{2} N^{2}}}\exp\left[\frac{-2\theta^{2}N}{1 + \xi^{2} N^{2}}\right].
\end{eqnarray}
%
\section{\label{sec:features}Features of the probability density}
%
Expressions (\ref{Ppm}) and (\ref{Pnn0_analytical}) for the probability density functions $P_{\pm}$ and $P_{0}$ give the probability $P(n_{+},n_{-},n_{0}) = P_{\pm}P_{0}$
of observing any given number of atoms in the three output ports (three atomic clouds) of an optical beamsplitter.
The function $P_{\pm}$ describes the probability of observing $n_{+}$ and $n_{-}$ atoms in the right and left clouds,
respectively, for a fixed number $n_{0} = N - (n_{+} + n_{-})$ atoms in the central cloud. This probability is the Gaussian distribution
(\ref{Ppm_Stirling}) with the average values of $n_{-}$ and $n_{+}$ given by
\begin{equation}\label{npnm_average}
    \langle n_{-} \rangle = \langle n_{+} \rangle = \frac{1}{2}(N-n_{0})
\end{equation}
and the standard deviations
\begin{equation}\label{npnm_std}
    \Delta n_{-} = \Delta n_{+} = \frac{1}{2}\sqrt{N-n_{0}}.
\end{equation}
The numbers of atoms in the right and left clouds are anti-correlated:
\begin{equation}\label{npnm_covariance}
    \mathrm{Cov}(n_{+}, n_{-}) = \langle n_{+}n_{-} \rangle - \langle n_{+} \rangle \langle n_{-} \rangle = -\frac{1}{4}(N-n_{0}).
\end{equation}
The probability to find $n_{0}$ atoms in the central cloud is given by the function $P_{0}(n_{0},\theta,\xi)$ Eq.~(\ref{Pnn0_analytical}). The dependence of this function on
its arguments is not trivial, so we start the analysis by evaluating the expectation value of the atoms on the central cloud $\langle n_{0} \rangle$ and the standard
deviation $\Delta n_{0}$.

The function $P_{0}$ is proportional to the modulus squared of the sum of two terms: $P_{0} \propto \left|e^{- \eta_{-}} + (-1)^{N-n_{0}}e^{-\eta_{+}}\right|^{2}$,
where $\eta_{\mp}$ are given by Eq.~(\ref{P0_exp_arguments}). The relative phase difference between them, as a function of $n_{0}$, changes rapidly due to the multiplier $(-1)^{n_{0}}$.
Thus, the interference terms can be neglected in calculating both the mean and the standard derivation:
\[\langle n_{0} \rangle \approx \frac{\sqrt{N}}{\sqrt{2\pi(1 + 4\xi^{2}N^{2})}}\int_{0}^{N}dn_{0}\sqrt{\frac{n_{0}}{N-n_{0}}}\left(e^{- 2\mathrm{Re}\eta_{-}} + e^{- 2\mathrm{Re}\eta_{+}}\right).\]
Evaluation of the above integral yields
\begin{equation}\label{n0_average}
    \langle n_{0} \rangle = \frac{N}{2}\left[1 + \exp\left(-2\xi^{2}N\right)\cos \theta\right].
\end{equation}

\begin{figure}
\includegraphics[width=8cm]{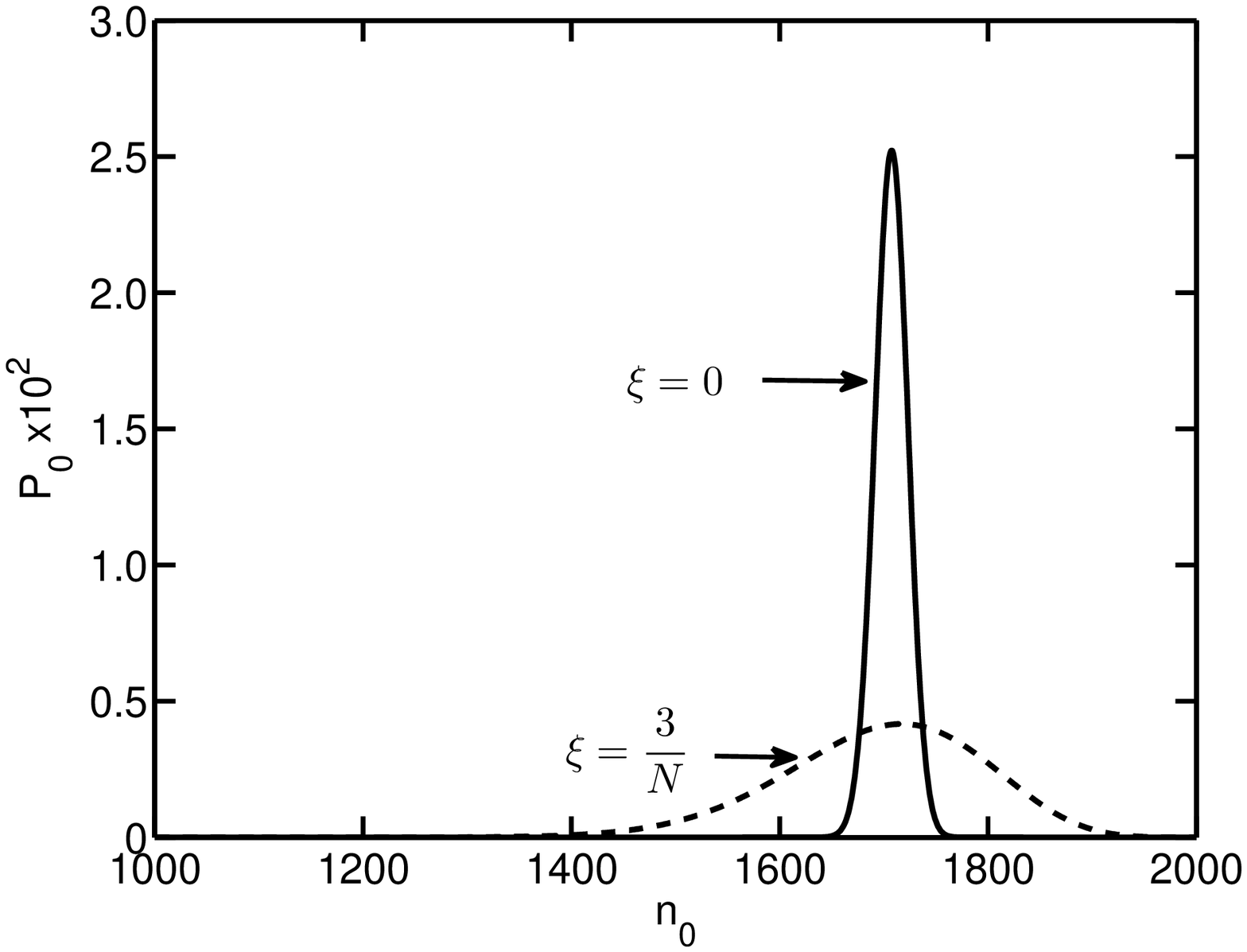}
\caption{\label{fig:P0_small_phi} Probability function $P_{0}$ versus $n_{0}$ for $\xi = 0$ and $\xi = 3/N$. For both curves, $\theta = \pi/4$
and $N = 2000$. Note that the abscissa axis range is from $n_{0} = 1000$ to $2000$.}
\end{figure}

Similarly, the standard deviation evaluates to
\begin{eqnarray}\label{n0_std}
    (\Delta n_{0})^{2} = \frac{N^2}{8}\left[1 - \exp\left(-4\xi^{2}N\right)\right]\left[1 - \exp\left(-4\xi^{2}N\right)\cos 2\theta\right].
\end{eqnarray}
To understand these results, we shall look at the dependence of the function $P_{0}$ on the number of atoms $n_{0}$ for different values of the strength of the interatomic interactions $\xi$.
At relatively small values of $\xi$ such that $\xi \ll 1/\sqrt{N}$, the term $\exp(-\eta_{-})$ in the expression (\ref{Pnn0_analytical})
for the probability dominates the second one. The probability $P_{0}$ is then a simple Gaussian
\begin{equation}
    P_{0} \propto  \exp\left[-\frac{2N\left(\arccos\sqrt{n_{0}/N} -\theta/2\right)^{2}}{1 + 4\xi^{2} N^{2}}\right]
\end{equation}

with the maximum located at $n_{0} = N\cos^{2}\theta/2$. This situation is shown in Fig.~\ref{fig:P0_small_phi}. The two curves in this figure are plots of the function
$P_{0}(n_{0})$ given by Eq.~(\ref{Pnn0_analytical})
versus $n_{0}$ for two difference values of the interatomic interaction strength $\xi$. Both curves correspond to the same value of the angle $\theta$.
The most noticeable feature of Fig.~\ref{fig:P0_small_phi} is the increase in the width of the probability distribution with $\xi$.
This behavior is explained by Eq.~(\ref{n0_average}), which in the limit  $\xi \ll 1/\sqrt{N}$ reduces to:
\begin{equation}
    \Delta n_{0} \approx \frac{\sqrt{N}}{2}\sin \theta \sqrt{1 + 4\xi^{2}N^{2}}.
\end{equation}
For very small values of $\xi$ ($\xi \ll 1/N$), the influence of the interatomic interactions on the operation of the beamsplitter is negligible.
The relative standard deviation of the number of atoms in the central cloud is inversely proportional to the square root of the total number of atoms in the system:
$\Delta n_{0}/N \propto 1/\sqrt{N}$.  For $1/N \ll \xi \ll 1/\sqrt{N}$, the width of the distribution linearly grows with the increase in $\xi$.

The mean value of $n_{0}$ for $\xi \ll 1/\sqrt{N}$ reasonably closely corresponds to the position of the peak. Equation (\ref{n0_average}) for $\langle n_{0} \rangle$ in this limit yields:
\begin{equation}
    \langle n_{0}\rangle \approx \frac{N}{2}\left(1 + \cos \theta \right).
\end{equation}
As is seen, $n_{0}$ depends on $\theta$ but not on $\xi$.

\begin{figure}
\includegraphics[width=8cm]{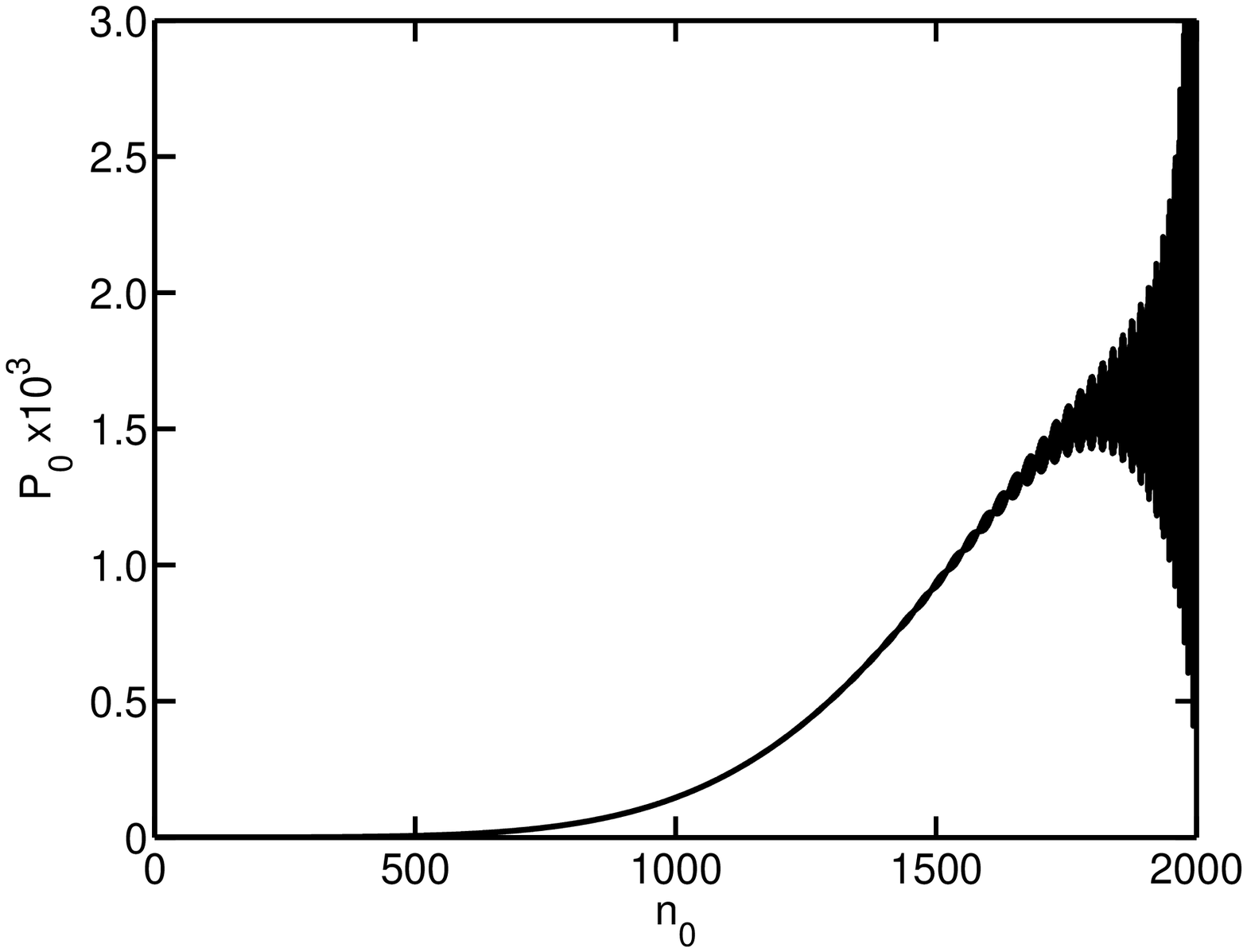}
\caption{\label{fig:P0_moderate_phi} Probability function $P_{0}$ versus $n_{0}$ for $\xi = 0.2/\sqrt{N}$, $\theta = \pi/4$
and $N = 2000$.}
\end{figure}

\begin{figure}
\includegraphics[width=8cm]{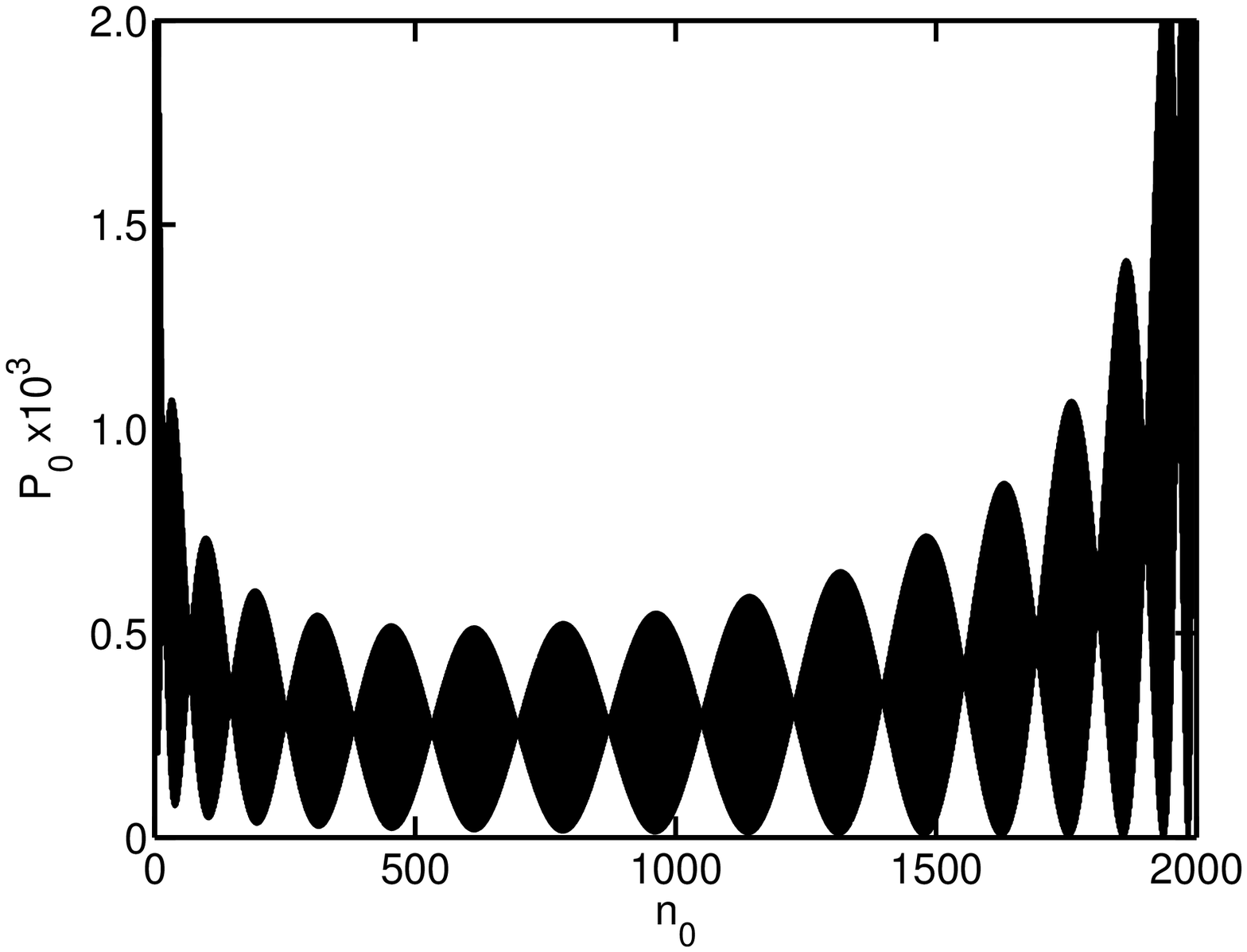}
\caption{\label{fig:P0_large_phi} Probability function $P_{0}$ versus $n_{0}$ for $\xi = 1/\sqrt{N}$, $\theta = \pi/4$
and $N = 2000$.}
\end{figure}

\begin{figure}
\includegraphics[width=8cm]{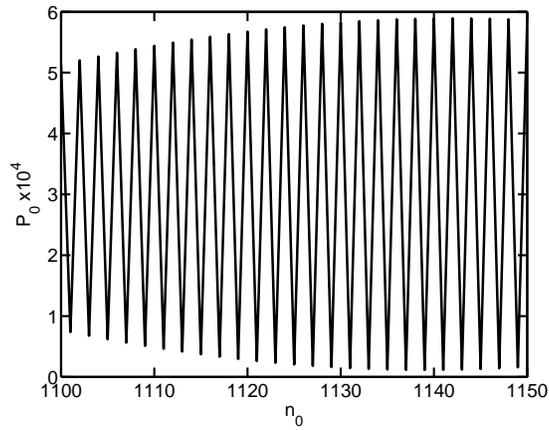}
\caption{\label{fig:P0_blowup} A blowup of a part of Fig.~(\ref{fig:P0_large_phi}) showing fast-scale spatial oscillations of the probability function.}
\end{figure}

For larger values of $\xi \approx 1\sqrt{N}$, the width of the probability distribution function $P_{0}$ becomes of the order of the total number of atoms in the system $N$.
The two terms $\exp(-\eta_{-})$ and $\exp(-\eta_{+})$ in Eq.~(\ref{Pnn0_analytical}) are now comparable in magnitude. The transition to this limit is shown by Fig.~\ref{fig:P0_moderate_phi}
and Fig.~\ref{fig:P0_large_phi}. Black regions not resolved in Figs.~\ref{fig:P0_moderate_phi} and \ref{fig:P0_large_phi} correspond to rapid spatial oscillations with the period $2$.
These oscillations are clearly seen in Fig.~\ref{fig:P0_blowup}, which shows part of Fig.~\ref{fig:P0_large_phi} for a narrow range of values of $n_{0}$. The oscillations are caused
by the interference between the two terms in Eq.~(\ref{Pnn0_analytical}). As the magnitude of $\xi$ approaches $1/\sqrt{N}$, these terms become comparable in magnitude. Because of the
nearly $\pi-$ phase change between the two terms every time $n_{0}$ changes by one due to the factor $(-1)^{n_{0}}$ , the two terms consecutively add
either in phase or out of phase when one steps through different values of $n_{0}$. Along with rapid spatial oscillations, both Fig.~\ref{fig:P0_moderate_phi} and Fig.~\ref{fig:P0_large_phi}
demonstrate oscillations of the envelopes at a much slower spatial rate which are more pronounced for larger values of the interaction strength. These oscillations are due to the fact that
the relative phase of the terms $\exp(-\eta_{-})$ and $\exp(-\eta_{+})$ in Eq.~(\ref{Pnn0_analytical}) changes with $n_{0}$. The nodes in Fig.~\ref{fig:P0_large_phi} correspond to the
value of this relative phase being equal to $0$ or a $\pi$ and an antinodes have the phase shifted by $\pm \pi/2$.

\begin{figure}
\includegraphics[width=8cm]{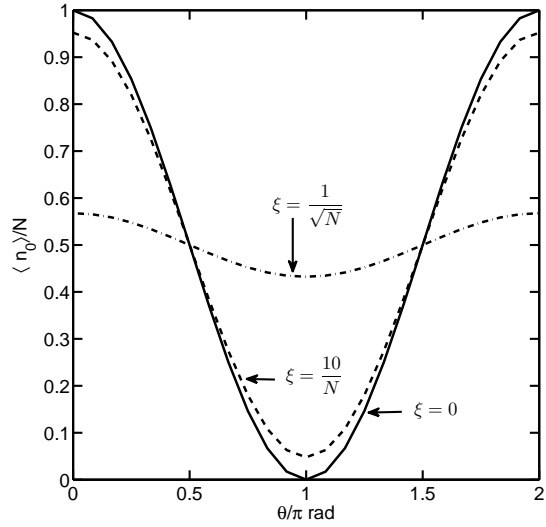}
\caption{\label{fig:avern0} Normalized mean value of the number of atoms in the central cloud $\langle n_{0} \rangle/N$ versus $\theta$ for $N = 2000$.}
\end{figure}

\begin{figure}
\includegraphics[width=8cm]{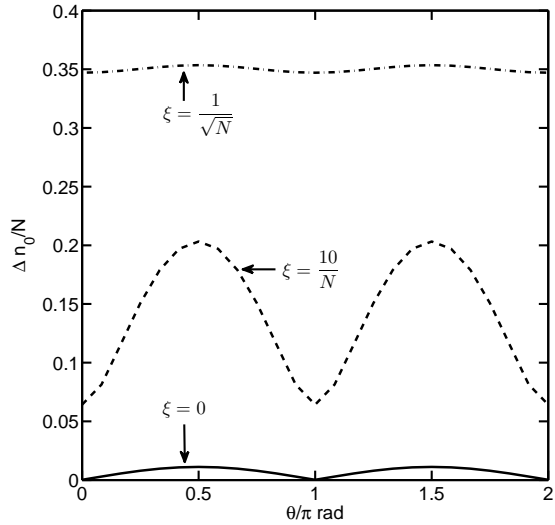}
\caption{\label{fig:stdn0} Normalized standard deviation $\Delta n_{0}/N$ versus $\theta$ for N = 2000.}
\end{figure}

Figs.~\ref{fig:P0_moderate_phi} and \ref{fig:P0_large_phi} indicate that the probability $P_{0}$ and, as a consequence, $\langle n_{0} \rangle$ and $\Delta n_{0}$, become
less sensitive to changes in the environment-introduced angle $\theta$. This fact is graphically illustrated by Figs.~\ref{fig:avern0} and \ref{fig:stdn0} showing the average value
of the number of atoms in the central cloud $\langle n_{0} \rangle$ and the standard deviation $\Delta n_{0}$ versus $\theta$ as given by Eqs.~(\ref{n0_average}) and (\ref{n0_std}), respectively.
Fig.~\ref{fig:avern0} demonstrates that increased interactomic interactions eventually lead to the loss of contrast of interference fringes. Additionally, larger interactomic interactions
cause larger shot-to-shot fluctuations in the number of atoms in each of the three output ports, as is seen from Fig.~\ref{fig:stdn0}.

\begin{figure}
\includegraphics[width=8cm]{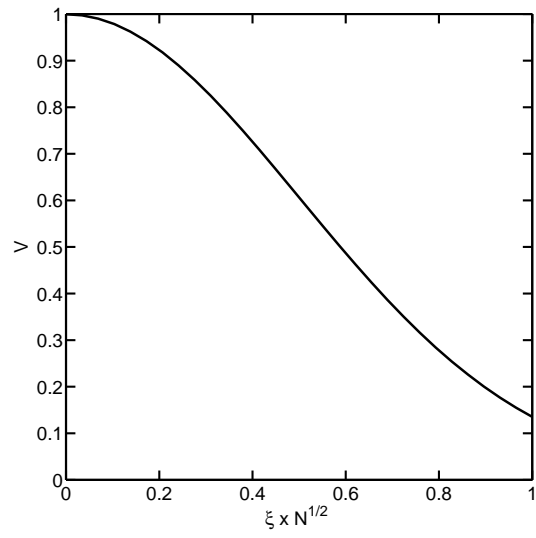}
\caption{\label{fig:contrast} Interference fringes contrast $V$ as a function of the strength of the interatomic interactions $\xi \sqrt{N}$.}
\end{figure}

The loss of contrast of the interference fringes can be quantified by writing Eq.~(\ref{n0_average}) as
\begin{equation}
\langle n_{0} \rangle = \frac{N}{2} \left(1 + V \cos \theta \right),
\end{equation}
where
\begin{equation}\label{contrast}
    V = \exp\left(-2\xi^{2}N\right)
\end{equation}
is the fringe contrast. Figure~\ref{fig:contrast} shows the fringe contrast V Eq.~(\ref{contrast}) as a function of $\xi$ and demonstrates that the values of $\xi$ approaching
$1/\sqrt{N}$ result in a washout of the interference fringes.
%
\section{\label{sec:discussion}Discussion}
%
The value of the accumulated nonlinear phase per atom due to interatomic interactions $\xi$ Eq.~(\ref{xi_definition})
depends on the volume of the BEC clouds (cf. Eq.~(\ref{g_definition})). Experiments \cite{wang05,garcia06,burke08,horikoshi06,horikoshi07,burke08}
to be discussed below, were conducted in parabolic traps with confining potentials of the form
\begin{equation}\label{parab_potential}
    V = \frac{M}{2}\left(\omega_{x}^{2}x^{2} + \omega_{y}^{2}y^{2} + \omega_{z}^{2}z^{2}\right).
\end{equation}
Density profiles of the moving clouds are well described by the Thomas-Fermi approximation
\begin{equation}\label{density_profile}
    n(\bm{r}) = |\psi_{\pm}|^{2} = \frac{15}{8\pi R_{x}R_{y}R_{z}}\left(1 - \frac{x^{2}}{R^{2}_{x}} - \frac{y^{2}}{R^{2}_{y}}- \frac{z^{2}}{R^{2}_{z}}\right)
\end{equation}
(recall that $\psi_{\pm}$ are normalized to one).

Immediately after the splitting pulses, the density profiles of the moving clouds are the same as that of the initial BEC
cloud containing $N$ atoms and being in equilibrium in the confining potential Eq.~(\ref{parab_potential}). After the splitting, each moving cloud contains
$N/2$ atoms. The repulsive nonlinearity is no more balanced by the confining potential and the radii of both clouds start to oscillate.
The maximum size of the oscillating clouds is the equilibrium size corresponding to $N$ atoms and the minimum size lies below the equilibrium
size corresponding to $N/2$ atoms. For estimates, we can take $R_{i}^{2}$ in Eq.~(\ref{density_profile}) to be given by equilibrium size of a cloud with
$N/2$ atoms: $R_{i}^{2} = 2\mu/M\omega_{i}^{2}$, where ~\cite{baym96,dalfovo99}
\begin{equation}\label{mu_N2}
    \mu = \frac{1}{4}\left(\frac{15}{\sqrt{2}\pi}\right)^{2/5}\left(N\frac{U_{0}}{\bar{a}^{3}}\right)^{2/5}\left(\hbar \bar{\omega}\right)^{3/5},
\end{equation}
$U_{0} = 4\pi\hbar^{2}a_{sc}/M$,  $\bar{\omega} = (\omega_{x}\omega_{y}\omega_{z})^{1/3}$ and $\bar{a} = \sqrt{\hbar/M\bar{\omega}}$.

Evaluation of the constant $g$ Eq.~(\ref{g_definition}) yields $g = (15U_{0})/(28 \pi R_{x}R_{y}R_{z})$.
The accumulated nonlinear phase per atom due to interatomic interactions $\xi$ Eq.~(\ref{xi_definition}) is then given by the expression
\begin{equation}\label{xi_estimate}
    \xi = \frac{1}{7}\left(30 \sqrt{2}\right)^{2/5}\left(\frac{a_{s}}{\bar{a}}\right)^{2/5}\bar{\omega}T N^{-3/5},
\end{equation}
where $T$ is the duration of the interferometric cycle.

The relative importance of interatomic interaction effects on the operation of the interferometer is determined by the parameter $P = \xi \sqrt{N}$:
\begin{equation}\label{P_estimate}
    P = 0.64\left(\frac{a_{s}}{\bar{a}}\right)^{2/5}\left(\bar{\omega}T\right)N^{-1/10}.
\end{equation}
Figure (\ref{fig:contrast}) shows that the contrast of the interference fringes decreases with the increase in $P$. The condition of good contrast can be
somewhat arbitrarily stated as $P < 1/2$ (for $P = 0.5$, $V = 0.6$).

Equation (\ref{P_estimate}) shows that $P \propto T \omega^{6/5} N^{-1/10}$. The dependence of $P$ on the total number of atoms in the BEC
clouds is very weak, and so this parameter is primarily dependent on the duration of the interferometric cycle and the averaged frequency of the trap.

Experiments by Wang et al~\cite{wang05} were conducted using the Michelson geometry. The BEC consisted of about $10^{5}$ $^{87}$Rb atoms~\cite{schwindt05}. The transverse and longitudinal
frequencies of the trap were $177\,$Hz and $5\,$Hz respectively.  The propagation time $T$ was up to $10\,$ms. For these parameters and the value of the scattering length
$a_{s} = 5.2\times 10^{-9}\,$m ~\cite{julienne97}, Eq.~(\ref{P_estimate}) yields $P \approx 1.6\times 10^{-2}$. Thus, the interatomic interactions were not limiting the visibility
of the interference fringes in these experiments.


A similar experiment was performed by  Garcia et al~\cite{garcia06,burke08} also in the geometry of a Michelson interferometer.
In Ref.~\cite{garcia06}, a BEC cloud of about $10^{4}$ $^{87}$Rb atoms has been produced in a trap with the frequencies of $6.0\,$Hz, $1.2\,$Hz and $3.0\,$Hz, respectively.
The interferometric time $T$ was about $40\,$ms. Using Eq.~(\ref{P_estimate}), we can evaluate the value of the parameter
$P$ in the experiment as $P \approx 10^{-2}$, which was too small to result in observed degradation of the contrast.  The loss of visibility in
the experiments \cite{garcia06} was attributed by the authors to spatial noise on the splitting beams and asymmetric splitting of the cloud due
to the condensate's residual motion when it was loaded into the trap. At longer tines, the loss of coherence might
have been caused by various noise sources. Similar results were reported in Ref.~\cite{burke08}, were the confinement frequencies were deliberately kept
weak making the atomic density and this the interatomic interaction effects small.

Horikoshi et al.~\cite{horikoshi06,horikoshi07} demonstrated a BEC Mach-Zehnder interferometer.
The number of atoms in Ref.~\cite{horikoshi07} was about $3 \times 10^{3}$ and the radial frequency of the trap was fixed at $60\,$Hz.
The experiments have been conducted for two different values of the axial frequencies and interrogation times $T$.
At an axial frequency of $\omega_{z} = 2\pi \times 17\,$Hz and the propagation time of the cloud about $T = 60\,$ms, the parameter
$P = \phi\sqrt{N}$ estimated using Eq.~(\ref{P_estimate}) turns out to be about $0.38$. For this value of $P$, Eq.~(\ref{contrast}) gives the value of the fringe
contrast about $70\%$. The experimental value is $30\%$ \cite{horikoshi07}. Similarly, for the axial frequency was  $10.29\,$Hz and the interferometric time $97\,$ms,
Eq.~(\ref{P_estimate}) gives the value of $P \approx 0.5\,$ corresponding to an estimated contrast of $58\%$. In this case no fringes were observed experimentally with a
about $40\%$ scatter of the data points. The authors of Ref.~\cite{horikoshi07} conjecture that vibrations could be the main source of the loss of fringes in their experiments.
The above estimates indicate that the interatomic interactions discussed in the present paper could be also partially responsible for the observed degradation
of the interference fringe.

\section{Acknowledgements}
This work was partially supported by the Defense Advanced Research Projects
Agency (Grant No. W911NF-04-1-0043). A.~A.~Z. thanks E.~Zhirova for helpful discussions.

\appendix
\section{Mach-Zehnder-type interferometer}

In a Mach-Zehnder-type cold-atom interferometer, the optical splitting $\pi/2$ pulse transforms a BEC
cloud at rest in a superposition of two clouds $\psi_{0}$ and $\psi_{+1}$. The first cloud is at rest and the second one is moving.
The clouds evolve during the time $T/2$ and are then subject to the action of a $\pi$ pulse. It stops the moving cloud and brings the one at rest
into motion, i.e., transforms the $\psi_{0}$ cloud into $\psi_{+1}$ and vice versa. After additional evolution time $T/2$,
the clouds are subject to a recombination $\pi/2$ pulse. After the recombination, both $\psi_{0}$ and $\psi_{+1}$ are in general populated.

Analysis of a Mach-Zehnder-type interferometer parallels that given in the paper for the Michelson-type interferometer and is somewhat
simpler because with the Mach-Zehnder interferometer there are only two output ports as opposed to three in the case of a Michelson-type interferometer.

Let $b^{\dagger}_{0}$ and $b^{\dagger}_{+1}$  be operators which create
an atom belonging to a cloud at rest and moving to the right, respectively.
The Hamiltonian is of the form (cf. Eq.~(\ref{hamiltonian_bbdagger})):

\begin{equation}\label{hamiltonian_bbdagger_MZ}
    H_{eff} = - \frac{W}{2}\left(\hat{n}_{+1} - \hat{n}_{0}\right) + g\left(\hat{n}_{+1}^{2} + \hat{n}_{0}^{2}\right),
\end{equation}
where the notations are the same as in Section \ref{sec:beamsplitter}.

The state vector of the system at the beginning of the interferometric cycle before the splitting pulse is given by Eq.~(\ref{psi_initial}).

The splitting/recombination $\pi/2$ pulse couples the operators $b_{1}$ and $b_{0}$
according to the rules:
\begin{eqnarray}\label{pi2_pulses_rules_MZ}
    &&b_{0} \rightarrow \frac{1}{\sqrt{2}}\left(b_{0} + i b_{+1}\right), \nonumber \\
    &&b_{+1} \rightarrow \frac{1}{\sqrt{2}}\left( i b_{0} - b_{+1}\right).
\end{eqnarray}
For the $\pi$ pulse, similarly, one has:
\begin{equation}\label{pi_pulses_rules_MZ}
    b_{0} \rightarrow i b_{+1},\; b_{+1} \rightarrow  -i b_{0}.
\end{equation}
Repeating steps of Section \ref{sec:beamsplitter}, we arrive at the following  expression for the state vector of the system after the recombination pulse:
\begin{eqnarray}\label{psi_after_recombination_MZ}
   &&|\Psi_{rec}\rangle = \frac{1}{2^{N}\sqrt{N!}}\sum_{n=0}^{N}\binom{N}{n}e^{ i\theta(n-N/2) - 2i\xi(n - N/2)^{2}} \nonumber \\
   &&\times \left(b^{\dagger}_{0} -i b^{\dagger}_{+1}\right)^{n}
   \left(b^{\dagger}_{0} + i b^{\dagger}_{+1}\right)^{N-n}|0\rangle
\end{eqnarray}
The probability to have after the recombination $n_{0}$ atoms at rest and $n_{+} = N - n_{0}$ atoms moving  is given by the modulus squared of the
probability amplitude $\langle n_{0},N - n_{0}|\Psi_{rec}\rangle$. Here
\begin{equation}\label{eigenstate_n0_np_MZ}
    |n_{0},n_{+}\rangle = \frac{\left(b_{0}^{\dagger}\right)^{n_{0}}}{\sqrt{n_{0}!}}\frac{\left(b_{+1}^{\dagger}\right)^{n_{+}}}{\sqrt{n_{+}!}}|0\rangle
\end{equation}
is the state with $n_{0}$ atoms at rest and $n_{+}$ atoms moving, respectively.

Repeating steps of Sec. \ref{sec:probability}, the probability $P(n_{0}) = |\langle n_{0},N-n_{0}|\Psi_{rec}\rangle|^{2}$
can be written as
\begin{equation}\label{Pn0_MZ}
    P(n_{0},\theta,\xi) = \binom{N}{n_{0}}\left|\Sigma(n_{0},\theta,\xi)\right|^{2},
\end{equation}
where $\Sigma$ is given by Eq.~(\ref{Sigma}) with the function $I(n,n_{0})$ in Eq.~(\ref{Sigma}) given by the relation (\ref{Inn0}).
The probability $P$ Eq.~(\ref{Pn0_MZ}) is thus completely identical to the probability  $P_{0}$ Eq.~(\ref{Pn0}) of Section \ref{sec:probability}.
All relations of Sections \ref{sec:evaluation} and \ref{sec:discussion} equally apply to the case of the Mach-Zehnder-type interferometer.
%
\newpage
\bibliographystyle{apsrev}

\newpage
%
%
%
%
%
%
%
%
%
\end{document}